\documentclass[final]{IEEEtran}
\ifCLASSINFOpdf
\else
\fi

\usepackage[ruled,linesnumbered,vlined]{algorithm2e}
\SetKw{KwBy}{by}
\usepackage{amsthm,amssymb,graphicx,multirow,amsmath,bm,color,amsfonts}
\usepackage[update,prepend]{epstopdf}
\usepackage{cite}
\usepackage{tabulary}
\usepackage{bbm} 
\usepackage{multirow}
\usepackage{comment}
\usepackage[caption=false]{subfig}
\usepackage[normalem]{ulem}

\makeatletter
\newcommand{\nosemic}{\renewcommand{\@endalgocfline}{\relax}}
\newcommand{\dosemic}{\renewcommand{\@endalgocfline}{\algocf@endline}}
\let\oldnl\nl
\newcommand{\nonl}{\renewcommand{\nl}{\let\nl\oldnl}}
\usepackage{makecell}

\usepackage{optidef}

\newcommand*{\hermconj}{^{\mathsf{H}}}

\def\nb0{{\mathbf{0}}}
\def\nb1{{\mathbf{1}}}

\def\sinc{{\rm sinc}}

\newtheorem{lemma}{Lemma}

\newtheorem{nrem}{Remark}

\def\figref#1{Fig.\,\ref{#1}}%

\allowdisplaybreaks 
\begin{document}
\pagenumbering{gobble}
\graphicspath{{./figures/}}
\title{
	Multi-User Precoding with Stacked Intelligent Metasurfaces via Neural Reparameterization
}
\author{
	Yongqiang Zhang, \IEEEmembership{Member,~IEEE,} and Qurrat-Ul-Ain Nadeem, \IEEEmembership{Senior Member,~IEEE} 
	\thanks{ Yongqiang Zhang and Qurrat-Ul-Ain Nadeem are with the Engineering Division, New York University Abu Dhabi (NYUAD), Abu Dhabi, UAE. 
		Qurrat-Ul-Ain Nadeem is also with the NYU Tandon School of Engineering, Brooklyn, NY 11201, USA, and  the NYUAD Wireless Center, NYUAD Research Institute, Abu Dhabi, UAE.
		(e-mails: \{yz12825,  qurrat.nadeem\}@nyu.edu)}
}

\maketitle

\begin{abstract}
	Stacked intelligent metasurfaces (SIMs) enable wave-domain signal processing, and provide a promising architecture for enhancing spectral efficiency in multi-user wireless communication systems.
	However, jointly optimizing transmit power and multi-layer phase shifts in SIM-assisted wireless systems under quality-of-service (QoS) constraints results in a large-scale nonconvex problem, for which conventional numerical optimization-based methods, such as alternating optimization (AO), suffer from high computational burden and limited scalability.
	In this paper, we develop a QoS constraint-aware neural reparameterization  framework for SIM-assisted downlink precoding.
	By embedding the cascaded wave-propagation structure of the SIM into a model-based differentiable architecture, both the phase shift configuration and transmit power allocation are reparameterized as parameters of a structured neural model that inherently satisfies feasibility constraints  of the original problem.
	The resulting formulation enables efficient optimization that does not rely on a channel-to-precoding mapping learned through offline training. Instead, it directly updates the reparameterized optimization variables to realize real-time precoding for any channel realization.
	Simulation results demonstrate that the proposed method achieves superior sum rate performance under QoS constraints, while significantly improving computational efficiency compared with conventional numerical optimization approaches.
\end{abstract}

\section{Introduction}
Sixth-generation (6G)  wireless networks will demand extreme spectral efficiency and precise spatial control, which can potentially be achieved through ultra-massive multiple-input multiple-output (MIMO) architectures \cite{Dang2020, Tataria2021, Wang2024}. 
However, conventional digital massive MIMO incurs prohibitive hardware costs, power consumption, and signal processing complexity due to its reliance on numerous active radio frequency (RF) chains and high-resolution data converters \cite{Heath2016}. 
To address these bottlenecks, single-layer metasurfaces, such as reconfigurable intelligent surfaces (RISs), have emerged as a promising paradigm to manipulate electromagnetic (EM) waves in the analog domain with significantly reduced hardware requirements \cite{DiRenzo2020, Venkat2020}. 
An artificial metasurface comprises a large number of low-cost and nearly passive meta-atoms, each of which can independently impose an adjustable phase shift on the incident EM waves \cite{Wu2021b}.
By dynamically tuning these phase shifts via a smart controller, such as a field programmable gate array (FPGA), the metasurface is capable of reshaping the wireless propagation environment to create favorable channel conditions \cite{Pan2022}.
Despite their energy efficiency benefits, these single-layer architectures are fundamentally constrained by limited degrees of freedom (DoF), which restricts their ability to execute high-resolution multi-user beamforming and complex interference mitigation \cite{Guo2020}.

Multi-layer diffractive metasurfaces were first developed in optics, where a stack of passive diffractive surfaces was shown to implement an all-optical physical neural network that executes inference at the speed of light rather than on a digital processor
\cite{Liu2022a, Lin2018}. 
Motivated by the limited DoF of single-layer metasurfaces and inspired by the notable success of multi-layer diffractive metasurface computing, the authors in \cite{An2023} introduced this layered architecture into wireless communications as the stacked intelligent metasurface (SIM).
A SIM-enabled base station~(BS) can perform precoding and combining as waves propagate through its successive SIM layers at the speed of
light, thereby alleviating the reliance on a large number of active RF chains and complex digital baseband processing, since precoding and combining are not implemented in the baseband but rather in the wave domain~\cite{An2023a}.
This wave-domain processing paradigm has inspired a variety of recent transmission designs~\cite{An2025,  Fang2026, Papazafeiropoulos2024a, Nadeem2024, 	Bahingayi2026, An2024a, Nerini2024, Perovic2025, Niu2025, Amiri2025}.
However, most of these works formulate the SIM precoding design problem as a high-dimensional nonconvex optimization problem and solve it using iterative numerical optimization algorithms, such as alternating optimization (AO).
As the number of cascaded layers and/or meta-atoms per layer grows to meet stringent performance requirements, these conventional numerical methods encounter severe scalability limitations and substantial computational overhead.
These limitations are discussed further in the related work section, and motivate the proposed neural reparameterization method introduced thereafter.

\subsection{Related Works}\label{subsec:related}

Recent literature has extensively explored the potential of SIM-enabled multi-user beamforming, primarily for sum rate maximization. 
In \cite{An2023}, a wave-based beamforming framework was introduced for SIM-assisted multi-user multiple-input single-output (MISO) downlink systems, where transmit power allocation and multi-layer phase shifts were jointly optimized from a continuous phase shift set via an AO algorithm. 
This line of work was further extended in \cite{An2025} to incorporate practical discrete phase shift constraints. 
Furthermore, a refined AO framework was developed in \cite{Bahingayi2026} to investigate hybrid precoding that jointly orchestrates digital beamforming and SIM phase shifts.
Beyond sum rate, alternative performance metrics have been investigated in \cite{Fang2026, Kavianinia2025, Perovic2025}. 
For instance, max-min rate and geometric mean rate optimization problems in SIM-assisted networks were studied in \cite{Fang2026} using a consensus alternating direction method of multipliers (ADMM) framework. 
The authors in \cite{Kavianinia2025} solved a secrecy rate maximization problem through AO combined with successive convex approximation (SCA), while energy efficiency for SIM-based broadcast MIMO systems was investigated in \cite{Perovic2025}. 
\textcolor{black}{Beyond narrowband systems, SIM-assisted wideband multicarrier transmission has been investigated in \cite{Li2026, Li2026a, Li2025a}, where the SIM phase shifts, which are common to all subcarriers, are optimized for point-to-point MIMO orthogonal frequency division multiplexing (OFDM) systems, multi-user MIMO OFDM systems with index modulation, and orthogonal frequency division multiple access (OFDMA) systems, respectively.}

However, the aforementioned studies predominantly treat SIM control as a high-dimensional nonconvex optimization problem, and focus on system capacity while overlooking the quality-of-service (QoS) constraint at the users.
To ensure reliable communication in interference-limited environments, it is imperative to transition from unconstrained objective maximization to QoS-aware precoding designs.
While such designs have been maturely established for single-layer RIS architectures \cite{Guo2020}, their application to multi-layer SIM systems is still at an early stage. 
For example, a SIM-enhanced integrated communication and computation framework was devised in \cite{Chen2025b} to minimize computation error while satisfying communication signal-to-interference-plus-noise ratio (SINR) requirements. 
In \cite{Huai2025}, by using the coati optimization algorithm and an iterative projected gradient ascent algorithm, the authors investigated a SIM-aided rate-splitting multiple access (RSMA) system, where the sum rate was maximized subject to user-specific rate thresholds.
\textcolor{black}{More recently, the authors in \cite{Zhang2026} proposed a variable decoupling-based block successive upper bound minimization (VD-BSUM) algorithm for the QoS-constrained sum rate maximization problem in SIM-assisted multi-user systems.}
Despite these efforts, efficiently solving the QoS-constrained sum rate maximization problem in SIM-assisted communication systems remains largely unexplored, given that \textcolor{black}{existing sum rate maximization designs} either ignore this constraint or rely on high-complexity iterative methods.

To alleviate the substantial computational burden of iterative numerical solvers in large-scale SIM systems, learning-based approaches have been introduced as promising alternatives. In \cite{Attiah2020}, a deep learning framework was developed to jointly design channel sensing and hybrid precoding in a massive MIMO system, while \cite{Yang2025} proposed a meta-learning-based algorithm to accelerate the optimization of phase shifts and power allocation in SIM-enabled systems. Furthermore, model-free techniques such as deep reinforcement learning (DRL) have been employed in \cite{Liu2025b} to optimize SIM parameters through interaction with the wireless environment. 
\textcolor{black}{Moreover, a multi-agent DRL framework was developed in \cite{Hua2026} to jointly optimize the SIM phase shifts, the operating modes of the access points, and the power allocation in a cell-free massive MIMO system where each access point is equipped with a SIM.}
Despite their promising performance, these approaches generally rely on extensive training procedures and require the output dimension of the learning model to scale with the considered precoding design space, which becomes challenging as the SIM dimension increases.
In \cite{Zayat2025}, the authors adopted a data-driven approach by embedding the SIM phase shifts as neurons of a neural network and training them as neural network parameters over a channel dataset, so as to minimize the symbol error rate of a point-to-point MIMO system with SIMs deployed at both the transmitter and receiver. 
Although such a parameterization alleviates the output-dimension burden of conventional learning-based designs, the framework in \cite{Zayat2025} remains subject to two key limitations. 
First, it optimizes only the SIM phase shifts, without jointly considering resource allocation such as power allocation, and its formulation cannot accommodate additional constraints commonly encountered in wireless network optimization, such as QoS requirements. 
Second, it learns a fixed phase shift configuration policy offline from a set of training channels and reuses it for new channel realizations, so its performance is prone to degradation once the actual channel deviates from the training distribution~\cite{Zhang2017Rethinking}.

\textcolor{black}{An alternative to such offline learning is neural reparameterization, in which the optimization variables are expressed through a neural network whose parameters are optimized directly for each problem instance, without requiring a training dataset. 
This concept was originally introduced in the context of structural optimization in~\cite{Hoyer2019}, where the design variables are generated by a neural network whose parameters are optimized for each design task without training data. A related idea was also explored in image restoration through the deep image prior framework in~\cite{Ulyanov2020}.}
\textcolor{black}{In wireless communications, a similar approach was proposed in \cite{Gao2022Online} and applied to sum-rate maximization in a single-layer RIS-assisted system without QoS constraints. Specifically, a dedicated neural network is optimized for each channel realization by treating the optimization variables as network parameters and the objective function as the loss function.
The network consists of two layers: a phase shift layer, whose parameters represent the phases of the RIS elements, followed by a beamforming layer that represents the digital precoding at the BS.
The authors showed that this approach can enable computationally efficient joint design of the BS precoding and the RIS phase shifts.}

\textcolor{black}{The above review indicates that several important challenges remain in SIM-assisted precoding design. First, conventional numerical optimization methods for SIM-assisted precoding generally involve computationally demanding iterative procedures, whose complexity increases with the SIM dimension. 
Second, efficient QoS-constrained sum-rate maximization in multi-layer SIM-assisted multi-user systems remains challenging.
Third, while learning-based approaches can reduce online computational complexity, they typically require offline training and may suffer from generalization issues. Although recent neural reparameterization approaches avoid offline training by optimizing the network parameters for each channel realization, their application to the joint phase-shift and power allocation design of multi-layer SIMs remains unexplored.
These observations motivate the present work, whose main contributions are outlined in the following subsection.}

\subsection{Contributions}\label{subsec:contributions}

\textcolor{black}{We consider a SIM-assisted multi-user MISO downlink system and investigate the joint optimization of transmit power allocation and multi-layer SIM phase shifts to maximize the sum rate subject to a total transmit power constraint and individual user QoS requirements.}
\textcolor{black}{To solve this problem, we develop a neural reparameterization framework that combines the per-channel optimization nature of conventional numerical methods with the neural-network representation of SIMs. Specifically, we recast the SIM phase shifts and transmit power allocation as neural network parameters that are optimized directly for each channel realization, leveraging mature parameter-update mechanisms whose computational efficiency has been demonstrated in large-scale neural network training. In this framework, the neural network is not used as a black-box predictor trained offline to learn a mapping from channel realizations to precoding configurations, but rather serves as a structured parameterization of the optimization variables. This approach aims to reduce the computational burden of SIM-assisted precoding optimization without requiring offline training.}

\textcolor{black}{The main contributions of this paper are summarized as follows:}

\begin{itemize}
	\item
	We formulate a QoS-constrained sum rate maximization problem for a SIM-assisted multi-user downlink MISO system, where a multiple-antenna BS serves multiple single-antenna users. 
	The goal is to jointly optimize the SIM phase response and the transmit power allocation across the users under a total power constraint and per-user minimum rate requirements.
	\item
		We establish a neural reparameterization framework that embeds the SIM phase shifts and transmit power allocation into a structured differentiable model whose forward pass mirrors the cascaded EM wave-propagation of the SIM. 
		Unlike black-box learning approaches that learn a channel-to-precoding mapping from data, the proposed framework recasts the original constrained problem as an equivalent optimization problem over neural network parameters.
	\item
	We further integrate the proposed neural reparameterization framework with a Lagrangian dual method to explicitly handle the QoS constraints. 
	By formulating a primal-dual optimization procedure, the neural parameters associated with the SIM phase shifts and power allocation are updated via a standard stochastic gradient optimizer, while the dual variables are adjusted through projected ascent.
	\item Through extensive simulations, we demonstrate that the proposed framework not only attains superior sum rate performance compared with the conventional numerical optimization schemes, but also exhibits remarkably enhanced computational efficiency.
	In particular, the proposed framework can achieve up to {a ten-fold reduction in computational time compared with} the AO baseline, thereby bringing wave-domain precoding in SIM-assisted transmission systems within reach of real-time operation.
\end{itemize}

The rest of this paper is organized as follows. Section \ref{sec:modelandproblem} introduces the SIM-assisted downlink system model and the QoS-constrained joint optimization problem for transmit power allocation and SIM phase shift design.
Section \ref{sec:framework} then develops a neural reparameterization framework by embedding the cascaded wave-propagation structure into a differentiable neural network architecture, along with the associated optimization procedure.
Simulation results in Section \ref{sec:results} demonstrate the effectiveness of the proposed approach.
Finally, Section \ref{sec:conclusion} concludes the paper.

\section{System Model and Problem Formulation}\label{sec:modelandproblem}

\begin{figure}[!htb]
	\center
	\includegraphics[width=1\columnwidth]{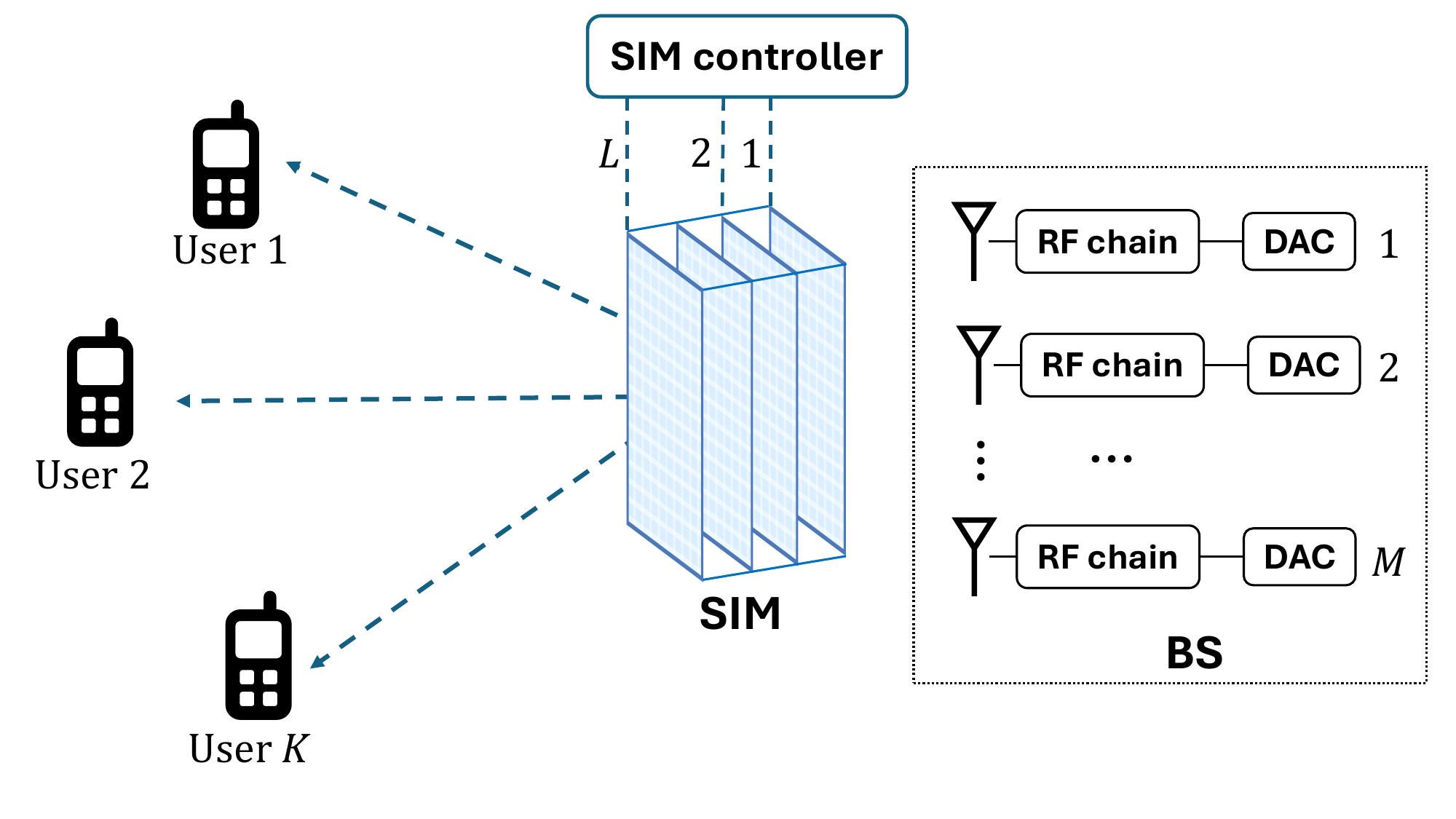}
	\caption{System model.}
	\label{fig:system_model}
\end{figure}

As depicted in \figref{fig:system_model}, we consider a downlink multi-user MISO communication system, where a SIM-enabled BS serves $K$ single-antenna users over a given frequency band.
The BS is equipped with $M$ transmit antennas, while the SIM comprises $L$ metasurface panels, each of which is equipped with $N$ programmable meta-atoms.
By configuring the phase response of these meta-atoms, the SIM regulates the EM waves transmitted from the BS prior to their propagation toward the users. 
A dedicated smart controller, such as a customized FPGA platform, is employed to enable independent phase adjustment for each meta-atom.

The phase shift matrix corresponding to the $l$-th layer is defined as
\begin{align}
	\bm{\Phi}_l = \mathrm{diag}\Big(a_1^l e^{j\theta_1^l}, \dots, a_n^l e^{j\theta_n^l},  \dots, a_N^l e^{j\theta_N^l}\Big) \in \mathbb{C}^{N \times N},
\end{align}
where $a_n^l \in \{0,1\}$ and $\theta_n^l \in [0, 2\pi)$ denote the amplitude control coefficient and the phase shift introduced by the $n$-th meta-atom in the $l$-th layer, respectively, 
$j$ denotes the imaginary unit satisfying $j^2 = -1$, and $\mathrm{diag}(x_1, \dots, x_N)$ constructs a diagonal matrix whose diagonal entries are given by the elements of the vector $\bm{x} = [x_1, \dots, x_N]^\top$, where $(\cdot)^\top$ represents the transpose operation.
To improve energy efficiency, meta-atoms are typically configured to operate in a transmission-maximizing mode by setting $a_n^l = 1$ for all \textcolor{black}{$1 \leq n \leq N$ and $1 \leq l \leq L$} \cite{Huang2019Reconfigurable,An2025, An2023a}.\footnote{\textcolor{black}{As in \cite{An2023a, Fang2026, Papazafeiropoulos2024a}, the considered SIM model assumes unit-amplitude transmission coefficients and continuous phase control, while neglecting insertion loss and mutual coupling among meta-atoms.} \textcolor{black}{The proposed framework in Section \ref{sec:framework} has the potential to accommodate more realistic SIM models by appropriately modifying its forward propagation structure.} \textcolor{black}{Incorporating these hardware nonidealities, including finite-resolution phase control, and evaluating their impact on the proposed optimization framework are left for future work.}}
For notational simplicity, the amplitude coefficient $a_n^l$ will be omitted in the subsequent analysis.

Based on the Rayleigh-Sommerfeld diffraction theory, the EM coupling between the $(l-1)$-th layer and the $l$-th SIM layer is characterized by the transmission matrix $\bm{W}_l \in \mathbb{C}^{N \times N}, \forall 2 \leq l \leq L$, whose $(n, n^\prime)$-th entry can be written as \cite{Lin2018}
\begin{align}\label{eq:tx_coeff}
	w^l_{n, n^\prime} = \frac{d_x d_y \cos \chi^l_{n, n^\prime} }{d^l_{n, n^\prime}} \left( \frac{1}{2\pi d^l_{n, n^\prime}} -\frac{j}{\lambda} \right)  \exp\left( {\frac{j 2\pi  d^l_{n, n^\prime} }{\lambda}}\right), 
\end{align} 
where $\lambda$ is the wavelength, $d^l_{n, n^\prime}$ is the transmission distance from the $n^\prime$-th meta-atom in the $(l-1)$-th layer to the $n$-th meta-atom in the $l$-th layer,  $\chi^l_{n, n^\prime}$ represents the angle between the propagation
direction from meta-atom $n^\prime$ to meta-atom $n$ and the normal
direction of the $(l-1)$-th metasurface layer, and the product of  $d_x$ and $d_y$ is the size of each meta-atom.

Similarly, to characterize the signal propagation from the BS to the SIM, let $d^1_{n, m}$ and $\chi^1_{n, m}$ denote the transmission distance and the corresponding propagation angle from the $m$-th transmit antenna to the $n$-th meta-atom on the first SIM layer, respectively. 
The $(n, m)$-th entry, denoted by $w^1_{n, m}$, of the transmission matrix $\bm{W}_1 \in \mathbb{C}^{N \times M}$ can be systematically evaluated by replacing $d^l_{n, n^\prime}$ and $\chi^l_{n, n^\prime}$ in \eqref{eq:tx_coeff} with $d^1_{n, m}$ and $\chi^1_{n, m}$. 
The transmission vector from the $m$-th transmit antenna to the first SIM layer is defined as the $m$-th column of $\bm{W}_1$, i.e., $\bm{w}_{m}^{1}  = [w_{1, m}^{1}, w_{2, m}^{1}, \dots, w_{N, m}^{1}]^\top \in \mathbb{C}^{N \times 1}$.

According to the Huygens-Fresnel principle, the wave propagation between adjacent metasurface layers can be modeled such that each meta-atom on the preceding layer transmits to all meta-atoms on the subsequent layer \cite{Papazafeiropoulos2024}. 
The incident field at each meta-atom is thus obtained by combining the signals radiated from all meta-atoms on the previous layer.
As a result, the SIM-enabled wave domain beamforming matrix can be modeled as
\begin{align}\label{eq:computeG}
	\bm{G} = \bm{\Phi}_{L}\bm{W}_{L}\bm{\Phi}_{L-1}\cdots \bm{\Phi}_{2}\bm{W}_{2}\bm{\Phi}_{1}.
\end{align}

We model the channels as quasi-static flat fading and assume that the channel state information (CSI) is perfectly known at the BS \cite{An2025, Papazafeiropoulos2024a, Bahingayi2026}.
Let $\bm{h}_k \in \mathbb{C}^{  N \times 1}$ denote the baseband equivalent channel from the last metasurface layer to the $k$-th user.
The channel is modeled as correlated Rayleigh fading
\begin{equation}
	\bm{h}_k \sim \mathcal{CN}\left(\mathbf{0}, \beta_k \bm{R}_k\right), \quad  1 \leq k \leq K,
\end{equation}
where $\beta_k $ denotes the large-scale fading coefficient accounting for the distance-dependent path loss between the SIM and user $k$, and $\bm{R}_k \in \mathbb{C}^{N \times N}$ is a positive semidefinite matrix characterizing the spatial correlation among the meta-atoms on the $L$-th layer. $\mathcal{CN}$ denotes the circularly symmetric complex Gaussian distribution.
Under the assumption of isotropic scattering with uniformly distributed multi-path components, the $(n, n')$-th entry of $\bm{R}_k$ is given by
\begin{equation}
	[\bm{R}_k]_{n, n'}  = \sinc \left( \frac{2  \varrho_{n,n'}}{\lambda} \right),
\end{equation}
where $\varrho_{n,n'}$ denotes the distance separating the $n$-th and the $n^\prime$-th meta-atoms on the last layer of the SIM.

Each user $k$ is assigned a data symbol  $s_k$, where $\{s_k\}_{k=1}^{K}$ are modeled as independent and identically distributed random variables with zero mean and unit variance.
The BS allocates a nonnegative transmit power $p_k$ to the $k$-th user, subject to the total power constraint
\begin{equation}\label{eq:powerallowance}
	\sum_{k=1}^{K} p_k \leq P_{\max},
\end{equation}
where $P_{\max}$ denotes the maximum available transmit power.

Similar to prior studies \cite{An2023, An2025, Fang2026, An2023a, Papazafeiropoulos2024a}, we focus exclusively on the wave-domain beamforming design enabled by the SIM.
Specifically, we consider that the SIM serves $K=M$ users using wave-domain beamforming without any digital precoding. 
After propagating through the SIM, the composite signal $y_k$ received at the $k$-th user is expressed as
\begin{align}\label{eq:recSig}
	y_k = \bm{h}_k\hermconj \bm{G} \bm{W}_1 \bm{P} \bm{s} + n_k,
\end{align}
where $\bm{P}  = \mathrm{diag}(\sqrt{p_1}, \dots, \sqrt{p_K})$, $\bm{s} = [s_1, \dots, s_K]^{\textcolor{black}{\top}}$, and  $n_k \sim \mathcal{CN}(0, \sigma_k^2)$ is the additive white Gaussian noise at the $k$-th user with noise power $\sigma_k^2$.
The operator $(\cdot){\hermconj}$ denotes the Hermitian transpose.

Treating the signals intended for the remaining users as interference, the corresponding SINR at the $k$-th user is expressed as
\begin{equation}\label{eq:SINR}
	\gamma_{k} =
	\frac{|\bm{h}_{k}\hermconj \bm{G}  \bm{w}_{k}^{1}|^{2} p_{k}}
	{\sum_{k' \neq k}^{K} |\bm{h}_{k}\hermconj \bm{G} \bm{w}_{k'}^{1}|^{2} p_{k'} + \sigma_{k}^{2}}.
\end{equation}

Consequently, the achievable data rate of the $k$-th user is given by
\begin{equation}\label{eq:user_rate}
	r_k =  \log_{2}(1 + \gamma_{k}).
\end{equation}

\subsection{Problem Formulation}
We aim to maximize the sum rate of all users under QoS constraints by jointly optimizing the transmit power at the BS and the wave-based beamforming at the SIM.
Define $\boldsymbol{\vartheta}^l = [\textcolor{black}{\theta_1^l, \theta_2^l, \dots, \theta_N^l}]^{\textcolor{black}{\top}} \in \mathbb{R}^{N\times 1}$.
Let $\bm{\Theta}=\{\boldsymbol{\vartheta}^1, \dots, \boldsymbol{\vartheta}^L\}$ denote the phase shift design policy. 
The optimization problem is formulated as 
\begin{subequations}\label{eq:optSumRate}
	\begin{align}
		(\textbf{OPT-1}):~&\underset{\bm{P},  \bm{\Theta}}{\max}~\, 
		\sum_{k=1}^{K}   r_k  \label{eq:obj_fun1}\\
		\text {s.t.} \qquad &  
		\sum_{k=1}^{K} p_k \leq P_{\max}, \label{eq:c1p1}\\
		&  r_k \geq R_{k}^{\min}, \quad \textcolor{black}{\forall k,} \label{eq:c2p1}\\
		& p_k \geq 0, \quad \textcolor{black}{\forall k,} \label{eq:c3p1}\\
		& \theta_n^l \in [0, 2\pi), \quad \textcolor{black}{\forall n, l,} \label{eq:c4p1}
	\end{align}
\end{subequations}
where \eqref{eq:c1p1} imposes the power budget constraint introduced earlier in \eqref{eq:powerallowance}.
\eqref{eq:c2p1} indicates that the achievable rate of user $k$ must not be less than the minimum data rate threshold $R_{k}^{\min}$.
\eqref{eq:c3p1} ensures the nonnegativity of the power allocated to each user, while \eqref{eq:c4p1} restricts the phase shift of each meta-atom to the interval $[0, 2\pi)$.

The optimization problem formulated in \eqref{eq:optSumRate} is highly nonconvex due to the cascaded EM propagation coefficients through the SIM layers, and the unit-modulus constraints imposed on the phase shifts. 
Conventional iterative optimization methods typically suffer from high computational complexity and poor scalability when the number of metasurface elements or layers increases. 
To overcome these limitations, we develop a neural reparameterization framework that embeds the deterministic wave-propagation model of the SIM into a structured differentiable neural network architecture.
This formulation enables the joint optimization of phase shifts and transmit power through efficient gradient-based updates of the neural network parameters directly, while preserving the underlying physical structure of the system. 
The proposed framework is detailed in the next section.

\section{Proposed Framework}\label{sec:framework}

The key idea of our proposed approach is to reformulate the optimization problem in  \eqref{eq:optSumRate} through a neural reparameterization of the SIM phase shifts and transmit power allocation. 
The precoded signal transmitted from the SIM is represented as the output of a structured neural network architecture whose forward propagation explicitly follows the deterministic EM wave-propagation model of the SIM.
In this formulation, the neural network does not act as a black-box predictor. 
Instead, it serves as a parametric realization of the SIM transmission model, and is therefore referred to as a neural reparameterization framework.
By constructing a loss function that is consistent with the objective of \eqref{eq:optSumRate}, the SIM phase shifts and the transmit power allocation can be optimized directly via backpropagation through the neural network parameters.

\subsection{Structured Forward Propagation}

In conventional neural networks, forward propagation maps an input vector to an output vector through a sequence of layers.
In a standard feedforward architecture, the input is processed by an input layer, followed by a sequence of hidden layers and an output layer \cite{goodfellow2016deep}.
The forward propagation through a typical layer  can be abstractly represented as 
\begin{align}\label{eq:forwardpass}
	\bm{y} = f\left(g \left(\bm{x}; \boldsymbol{\phi} \right) + \bm{b}\right),
\end{align}
where $\bm{x}$ is the input vector of the layer, $\bm{y}$ represents the output vector of the layer, $g(\cdot ; \boldsymbol{\phi} )$ is a parameterized transformation with parameter set $\boldsymbol{\phi}$, $\bm{b}$ denotes the bias vector, and $f(\cdot)$ corresponds to the activation function.

\begin{figure}[!htb]
	\center
	\includegraphics[width=1\columnwidth]{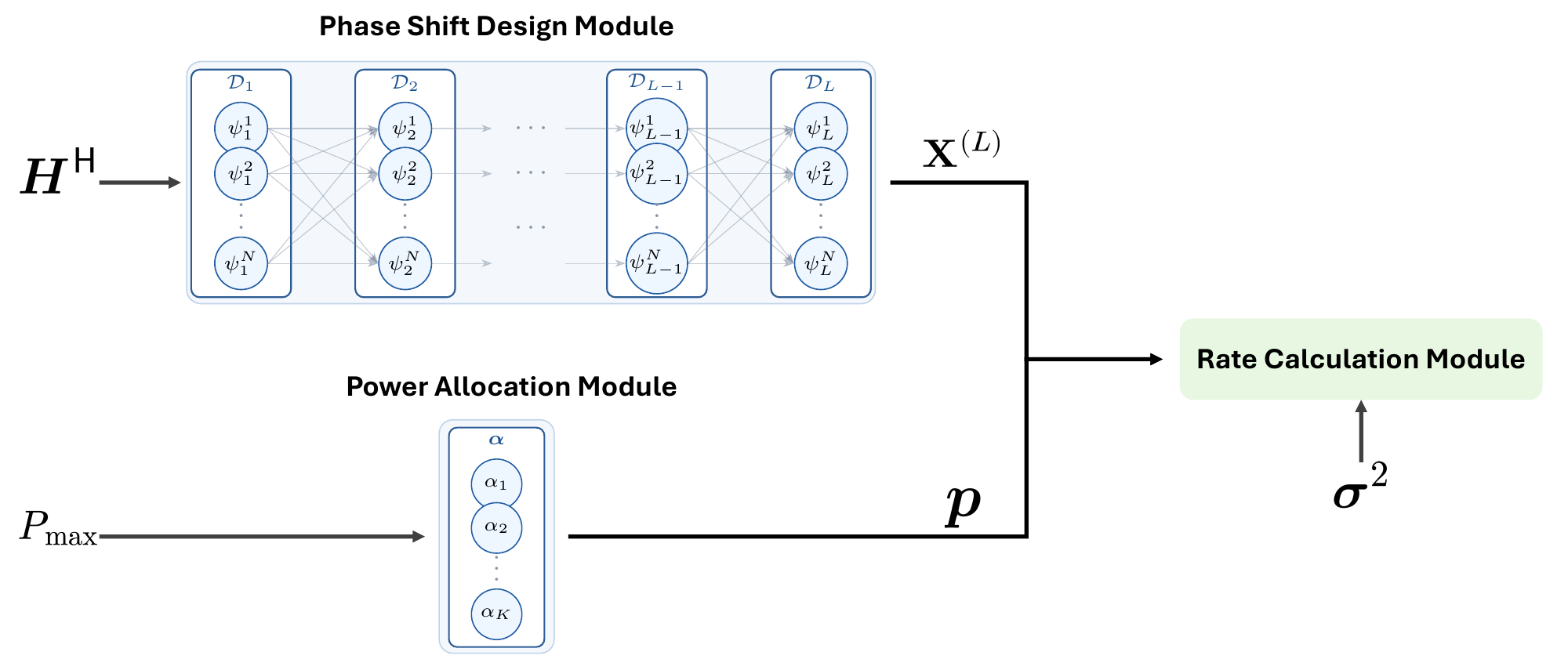}
	\caption{Illustration of the structured neural network $\mathcal{F}$.}
	\label{fig:prop_module}
\end{figure}

To solve the formulated problem (\textbf{OPT-1}), we construct a structured neural network $\mathcal{F}$, as shown in \figref{fig:prop_module}.
The proposed architecture consists of three modules: a phase shift design module, a power allocation module, and a rate calculation module.
In what follows, $\bm{H}=[\bm{h}_1,\dots,\bm{h}_K]\in\mathbb{C}^{N\times K}$ denotes the aggregated downlink channel matrix and $\boldsymbol{\sigma}^2\triangleq[\sigma_1^2,\dots,\sigma_K^2]^{\textcolor{black}{\top}}$ represents the users' noise power profile.
The phase shift design module operates on the Hermitian transpose of the aggregated channel matrix, i.e., $\bm{H}\hermconj\in\mathbb{C}^{K\times N}$, and is implemented as an $L$-layer network that mirrors the $L$ stacked metasurface layers of the SIM. 
The associated learnable \textcolor{black}{neural parameters are} denoted by $\boldsymbol{\psi}$, \textcolor{black}{from which} the optimization variables $\bm{\Theta}$ in (\textbf{OPT-1}) \textcolor{black}{are obtained}.
Given the total power budget $P_{\max}$, the power allocation module determines the power allocation policy $\bm{P}$, which is defined in \eqref{eq:recSig}.
This decision is parameterized by the learnable vector $\boldsymbol{\alpha}$.
Finally, the rate calculation module contains no learnable parameters.
It combines the outputs of the first two modules with the noise power vector $\boldsymbol{\sigma}^2$ to compute the per-user rate vector $\bm{r}\triangleq[r_1,\dots,r_K]^{\textcolor{black}{\top}}\in \mathbb{R}^K$ according to \eqref{eq:SINR}-\eqref{eq:user_rate}.
In summary, the proposed neural reparameterization network can be expressed as
\begin{align}\label{eq:pann}
	\bm{r} = \mathcal{F}(\bm{H}\hermconj,P_{\max},\boldsymbol{\sigma}^2; \boldsymbol{\psi}, \boldsymbol{\alpha}).
\end{align}

\subsubsection{Phase Shift Design Module}

The phase shift design module is constructed to explicitly follow the layer-wise physical propagation model of the SIM in \eqref{eq:computeG}. 
Given the input channel matrix $\bm{H}\hermconj$, the module performs $L$ successive structured transformations, where the depth of the neural architecture matches the number of metasurface layers. 
Each network layer corresponds to one physical SIM layer, thereby preserving the intrinsic EM propagation structure.

Due to the multiplication order implied by \eqref{eq:recSig} and \eqref{eq:computeG}, the neural network layer indexing is arranged such that the first transformation in the phase shift design module corresponds to the $L$-th layer of the SIM. 
Let $\boldsymbol{\psi} = [\boldsymbol{\psi}_1^\top,\dots,\boldsymbol{\psi}_L^\top]^\top$, where $\boldsymbol{\psi}_\ell = [\psi_\ell^1,\dots,\psi_\ell^N]^\top \in \mathbb{R}^N$ contains the parameters of the $\ell$-th network layer.
To clearly distinguish the learnable neural parameters from the physical phase shifts, we associate each entry $\psi_\ell^n$ with the phase shift of the $n$-th meta-atom in the $(L+1-\ell)$-th SIM layer through an element-wise mapping
\begin{align}\label{eq:phase_map}
\textcolor{black}{	\theta_n^{L+1-\ell} = \mathcal{Q}(\psi_\ell^n) \in [0,2\pi),}
\end{align}
where $\mathcal{Q}(\cdot)$ enforces the phase constraint in \eqref{eq:c4p1}.
\textcolor{black}{Specifically, $\mathcal{Q}(\cdot)$ is the modulo operation $\mathcal{Q}(\boldsymbol{\psi}_\ell) \triangleq \boldsymbol{\psi}_\ell - 2\pi \lfloor \boldsymbol{\psi}_\ell/(2\pi) \rfloor$, where the floor operator $\lfloor \cdot \rfloor$ is applied element-wise.}
\textcolor{black}{This operation is applied only as a post-processing step to the final values of $\boldsymbol{\psi}_\ell$, and it is not differentiated during the update of the neural parameters.}
\textcolor{black}{Instead, since $e^{j\mathcal{Q}(\boldsymbol{\psi}_\ell)} = e^{j\boldsymbol{\psi}_\ell}$, the phase shift matrices defined below are computed during this update as $\mathrm{diag}(e^{j\boldsymbol{\psi}_\ell})$, which is differentiable with respect to $\boldsymbol{\psi}_\ell$.}

For the first layer ($\ell=1$), the corresponding diagonal phase shift matrix is defined as
\begin{align}
	\mathcal{D}_1(\boldsymbol{\psi}_1)
	=
	\mathrm{diag}
	\big(
	\exp(j\textcolor{black}{\psi_1^1}),
	\dots,
	\exp(j\textcolor{black}{\psi_1^N})
	\big)
	\in \mathbb{C}^{N \times N}.
\end{align}

Since the structured transformation  does not involve a bias term and adopts the identity activation function, the output of the first layer simplifies from the general form in \eqref{eq:forwardpass} to
\begin{align}
	\bm{X}^{(1)}
	=
	\mathcal{G}_1^{(\theta)}(\bm{H}\hermconj; \boldsymbol{\psi}_1)
	=
	\bm{H}\hermconj
	\mathcal{D}_1(\boldsymbol{\psi}_1)
	\bm{W}_L.
\end{align}

For the subsequent layers, $2 \le \ell \le L$, the transformations follow the same structured recursion. 
Specifically, the diagonal phase shift matrix associated with the $\ell$-th network layer is given by
\begin{align}
	\mathcal{D}_\ell(\boldsymbol{\psi}_\ell)
	=
	\mathrm{diag}
	\big(
	\exp(j\textcolor{black}{\psi_\ell^1}),
	\dots,
	\exp(j\textcolor{black}{\psi_\ell^N})
	\big).
\end{align}

The output of the $\ell$-th layer is recursively expressed as
\begin{align}
	\bm{X}^{(\ell)}
	&=
	\mathcal{G}_\ell^{(\theta)}(\bm{X}^{(\ell-1)}; \boldsymbol{\psi}_\ell) \notag\\
	&=
	\bm{X}^{(\ell-1)}
	\mathcal{D}_\ell(\boldsymbol{\psi}_\ell)
	\bm{W}_{L+1-\ell},
	\quad 2 \le \ell \le L.
\end{align}

After $L$ successive transformations, the final output becomes
\begin{align}
	\bm{X}^{(L)}
	=
	\bm{H}\hermconj
	\left(
	\prod_{\ell=1}^{L}
	\mathcal{D}_\ell(\boldsymbol{\psi}_\ell)
	\bm{W}_{L+1-\ell}
	\right)
	=
	\bm{H}\hermconj \bm{G}  \bm{W}_1.
\end{align}

\subsubsection{Power Allocation Module} 
In parallel with the phase shift design module, we design a differentiable power allocation module that maps the transmit power budget $P_{\max}$ at the BS to a feasible user-wise power vector $\bm{p} = [p_1, p_2, \dots, p_K]^\top$.
To maintain structural consistency with the general layer-wise propagation model in \eqref{eq:forwardpass}, this module is formalized as a specialized parameterized layer.

Specifically, the learnable parameter of this module is $\boldsymbol{\alpha} = [\alpha_1,\dots,\alpha_K]^\top \in \mathbb{R}^K$.
The power vector $\bm{p}$ is generated via a structured transformation $\mathcal{G}^{(p)}(P_{\max};\boldsymbol{\alpha})$.
Since the objective of this module is to strictly enforce the physical power constraint without introducing unnecessary affine shifts, the bias term is set to $\bm{b}=\bm{0}$.
The forward propagation of $\mathcal{G}^{(p)}$ is therefore expressed as
\begin{align}\label{eq:power_forward}
	\bm{p} = \mathcal{G}^{(p)}(P_{\max}; \boldsymbol{\alpha})
	= P_{\max} f_{p}(\boldsymbol{\alpha}),
\end{align}
where $f_{p}(\boldsymbol{\alpha})$ is implemented using the \FuncSty{softmax} activation function, which produces a normalized power splitting vector $\boldsymbol{\pi} = [\pi_1,\dots,\pi_K]^\top \in \mathbb{R}^K$ \cite{Bishop2006}.
The $k$-th element of $\boldsymbol{\pi}=f_{p}(\boldsymbol{\alpha})$ is given by
\begin{align}\label{eq:softmax}
	\pi_k = \frac{\exp(\alpha_k)}{\sum_{i=1}^{K} \exp(\alpha_i)}.
\end{align}

By the construction given in \eqref{eq:power_forward} and \eqref{eq:softmax}, the power allocation module parameterizes the power vector such that $p_k = \pi_k P_{\max}, \forall 1 \leq k \leq K$, which automatically satisfies the feasibility conditions in \eqref{eq:c1p1} and \eqref{eq:c3p1}.

\subsubsection{Rate Calculation Module} 
The rate calculation module computes the achievable downlink rate for each user based on the noise power vector $\boldsymbol{\sigma}^2$, the power vector $\textcolor{black}{\bm{p}}$ generated by the power allocation module, and the composite channel matrix $\textcolor{black}{\bm{X}^{(L)}}$ obtained from the phase shift design module.
It computes the user rate vector $\bm{r}$ by using \eqref{eq:SINR} and \eqref{eq:user_rate}.
It is worth noting that this module does not introduce any additional trainable parameters since the rate calculation module only performs deterministic computations dictated by the signal transmission model.
Formally, let $\bm{B} = \bm{X}^{(L)} \odot\tilde{\bm{X}}^{(L)} \in \mathbb{R}^{K \times K}$ denote the effective power gain matrix, where $\tilde{\bm{X}}^{(L)} $ denotes the element-wise complex conjugate of $\bm{X}^{(L)}$ and $\odot$ represents the Hadamard product. 
Consequently, the vector of achievable downlink data rates can be expressed as
\begin{align}\label{eq:rate_vector}
	\bm{r} = \log_2 \left( \mathbf{1}_K + \frac{(\bm{B} \odot \bm{I}_K) \bm{p}}{(\bm{B} - \bm{B} \odot \bm{I}_K) \bm{p} + \bm{\sigma}^2} \right),
\end{align}
where $\bm{I}_K$ is the $K \times K$ identity matrix and $\mathbf{1}_K$ is an all-ones vector of length $K$.

In summary, the proposed structured forward propagation architecture seamlessly cascades three functional modules to map the initial system states directly to the achievable user rate vector $\bm{r}$. These initial states comprise the aggregated channel matrix $\bm{H}$, the total power budget $P_{\max}$, and the noise power profile $\boldsymbol{\sigma}^2$. 
While the phase shift design and power allocation modules encapsulate the network's learnable variables, $\boldsymbol{\psi}$ and $\boldsymbol{\alpha}$, the rate calculation module acts as a parameter-free output layer.
Crucially, all intermediate mathematical operations, including the softmax-based power mapping, the cascaded phase shift transformations, and the final rate calculation, are analytically differentiable. 
From an optimization perspective, this fully differentiable design is instrumental for maximizing the system objective, as it enables the systematic update of neural network parameters via the backward propagation mechanism detailed in the subsequent subsection.

\subsection{Sum Rate-Driven Backward Propagation}\label{subsec:backward}
As established in the previous subsection, the neural network parameters $\boldsymbol{\psi}$ and $\boldsymbol{\alpha}$ \textcolor{black}{determine} the SIM phase shifts $\bm{\Theta}$ and the transmit power allocation $\bm{P}$ \textcolor{black}{through \eqref{eq:phase_map} and \eqref{eq:power_forward}}.
 Consequently, the original precoding problem can be addressed by updating the corresponding neural network parameters to maximize the sum rate under the imposed constraints.
 In what follows, we detail how these updates are carried out.

In the neural network paradigm, while forward propagation evaluates the output through a sequence of transformations, backward propagation constitutes the fundamental mechanism for training by systematically updating the learnable parameters \cite{Rumelhart1986}. 
The primary objective is to optimize a predefined scalar loss function, denoted by $\mathcal{L}$, which quantifies the discrepancy between the current network output and the desired output.
Backward propagation computes the exact gradient of $\mathcal{L}$ with respect to the network parameters via recursive application of the multivariate chain rule. 
The essence of this recursion lies in the backward flow of error signals across the hierarchical architecture: the gradient with respect to the output of the $\ell$-th layer is determined from the sensitivity propagated from the subsequent $(\ell+1)$-th layer. Formally, let $\bm{y}^{(\ell)}$ and $\textcolor{black}{\boldsymbol{\phi}^{(\ell)}}$ denote the output vector and the learnable variables of the $\ell$-th layer, respectively. Once the gradient $\nabla_{\bm{y}^{(\ell)}} \mathcal{L}$ is obtained, the gradient with respect to the parameters of the $\ell$-th layer is expressed as
\begin{align}
	\nabla_{\textcolor{black}{\boldsymbol{\phi}^{(\ell)}}} \mathcal{L} = \left( \frac{\partial \bm{y}^{(\ell)}}{\partial \textcolor{black}{\boldsymbol{\phi}^{(\ell)}}} \right)\hermconj \nabla_{\bm{y}^{(\ell)}} \mathcal{L}. 
\end{align}
This recursive formulation serves as the computational cornerstone of modern learning systems, as it systematically reuses intermediate sensitivities to achieve efficient gradient evaluation, thereby enabling computationally efficient solutions to large-scale optimization problems within contemporary automatic differentiation frameworks such as PyTorch \cite{pytorch}.

To solve the constrained sum rate maximization problem formulated in (\textbf{OPT-1}), we first analyze the physical constraints in the context of the proposed forward propagation architecture. 
Notably, the constraints regarding the maximum total transmit power in \eqref{eq:c1p1} and the nonnegativity of the power allocation in \eqref{eq:c3p1} can be inherently satisfied by the \FuncSty{softmax} activation mappings integrated within our proposed neural reparameterization architecture.
Furthermore, the physical bounds governing the SIM phase shifts, as specified in the phase constraints, can be effectively addressed by applying a modulo post-processing operation to the obtained solutions.
The final value of \textcolor{black}{$\theta_n^l$} is calculated from \textcolor{black}{\eqref{eq:phase_map}} through the following transformation
\begin{equation}\label{eq:phase_forward}
	\textcolor{black}{\theta_n^l = \mathcal{Q}\big(\psi_{L+1-l}^n\big)} = \psi_{L+1-l}^n - 2\pi \left\lfloor \frac{\psi_{L+1-l}^n}{2\pi} \right\rfloor.
\end{equation}

Consequently, the user-specific QoS requirements in \eqref{eq:c2p1} remain the sole inequality constraints that require explicit mathematical treatment.
To tackle this, we introduce a set of non-negative Lagrangian multipliers $\boldsymbol{\mu} = [\mu_1, \dots, \mu_K]^\top \succeq \bm{0}$ associated with the QoS constraints, and transform the primal problem (\textbf{OPT-1}) into the following Lagrangian dual problem
\begin{align}\label{eq:opt2}
	(\textbf{OPT-2}): \quad \underset{\boldsymbol{\mu} \succeq \bm{0}}{\min} \,\, \underset{\boldsymbol{\psi}, \boldsymbol{\alpha}}{\max} \,\, \sum_{k=1}^{K} & r_k(\boldsymbol{\psi}, \boldsymbol{\alpha}) \notag \\
	& - \sum_{k=1}^{K} \mu_k \Big(R_k^{\min} - r_k(\boldsymbol{\psi}, \boldsymbol{\alpha})\Big),
\end{align}
where the dual variable $\mu_k$ can be interpreted as the price assigned to violating the QoS requirement of user $k$. 
When $r_k$ falls below its target, the penalty term decreases the overall Lagrangian objective in the primal maximization step, thereby encouraging subsequent parameter updates to improve the achievable downlink data rate of that user.

In problem (\textbf{OPT-2}), the objective function serves exactly as the basis for the unsupervised loss function for our backward propagation. 
To align with the standard neural network training paradigm where the goal is to minimize the loss function, we reformulate the sum rate maximization as the minimization of its negative counterpart. 
During the forward and backward propagation of the neural network, the dual variables $\boldsymbol{\mu}$ are treated as fixed penalty weights. 
Thus, the resulting loss function used for updating the primal network parameters  $\boldsymbol{\psi}$ and $\boldsymbol{\alpha}$ is constructed as follows
\begin{align}\label{eq:loss_PD}
	\mathcal{L}(\boldsymbol{\psi}, \boldsymbol{\alpha}, \boldsymbol{\mu})  =
	- \sum_{k=1}^{K} & r_k(\boldsymbol{\psi}, \boldsymbol{\alpha})
	+  
	\sum_{k=1}^{K}
	\mu_k \big(R_k^{\min} - r_k(\boldsymbol{\psi}, \boldsymbol{\alpha})\big).
\end{align}

\begin{algorithm}[!htbp]
	\caption{Proposed Algorithm for QoS-Constrained SIM-Assisted Precoding}
	\label{alg:proposed}
	\SetAlgoLined
	\KwIn{ \\
		\quad $\bm{H}$: CSI \\	
		\quad $P_{\max}$: maximum transmit power \\
		\quad ${\{R_k^{\min}\}_{k=1}^K}$: minimum data rate thresholds \\ 
		\quad $T$: maximum number of iterations
		
	} 
	\KwOut{ \\
		\quad $\bm{P}^*$:  power allocation policy \\
		\quad $\bm{\Theta}^*$:  phase shift design policy} 
	\BlankLine
	{Randomly initialize $\bm{\Theta}^{(0)}$ and set $\bm{P}^{(0)}$ to the equal power allocation}\;
	Initialize parameters $\boldsymbol{\psi}$ and $\boldsymbol{\alpha}$ of  $\mathcal{F}$ according to  $\bm{\Theta}^{(0)}$ and  $\bm{P}^{(0)}$\;
	Initialize dual variables $\boldsymbol{\mu} = [\mu_1, \dots, \mu_K]^\top = \bm{0}$\;
	\BlankLine
	\For{$t = {0, \dots, T-1}$}{
		\tcp{Forward Propagation}
		Compute the user rate vector $\bm{r}^{(t)} = \mathcal{F}(\bm{H}\hermconj, P_{\max}, \boldsymbol{\sigma}^2; \boldsymbol{\psi}, \boldsymbol{\alpha})$\;
		
		\BlankLine
		\tcp{Backward Propagation}
		Compute the loss function according to \eqref{eq:loss_PD}\;
		Update parameters $\boldsymbol{\psi}$ and $\boldsymbol{\alpha}$ using a stochastic gradient-based optimizer\;
		Update $\boldsymbol{\mu}$ according to \eqref{eq:updates}\;
	}
	\BlankLine
	\tcp{Post-processing}
	Compute the optimized $\bm{P}^*$ from $\boldsymbol{\alpha}$ according to \eqref{eq:power_forward}\;	
	Compute the optimized $\bm{\Theta}^*$ from $\boldsymbol{\psi}$ according to \eqref{eq:phase_forward}\;
	\KwRet{$\{\bm{P}^*, \bm{\Theta}^* \}$}
\end{algorithm}

To solve the minimax problem, after the neural parameters $\boldsymbol{\psi}$ and $\boldsymbol{\alpha}$ are optimized to minimize the loss function \eqref{eq:loss_PD} via backpropagation, the dual variables are then updated via projected gradient ascent to adjust the penalty for QoS violations.
The explicit update rule for the Lagrangian multiplier associated with user $k$ at the $t$-th iteration can be formulated as
\begin{align} \label{eq:updates}
	{\mu}_k & \leftarrow \left[ {\mu}_k + \textcolor{black}{\eta_{\mu}^{(t)}} \big(R_k^{\min} - r_k(\boldsymbol{\psi}, \boldsymbol{\alpha})\big) \right]^+,
\end{align}
where $t$ is the iteration index and \textcolor{black}{$\eta_{\mu}^{(t)}$ is a positive step size satisfying $\sum_{t}\eta_{\mu}^{(t)}=\infty$}. The projection operator $[\cdot]^+\triangleq \max(0, \cdot)$ ensures the nonnegativity of the Lagrangian multipliers.

For brevity, the proposed neural reparameterization algorithm for QoS-constrained SIM-assisted precoding is referred to as NR-QoS. 
The pseudocode of NR-QoS is summarized in Algorithm \ref{alg:proposed}.
At each iteration, the algorithm systematically updates the primal neural network parameters and the dual Lagrangian multipliers to jointly optimize the SIM phase shifts and users power allocation while accounting for the QoS constraints.

\begin{nrem}\label{remark:direct}
	\textcolor{black}{In Algorithm~\ref{alg:proposed}, the learnable neural parameters $\boldsymbol{\psi}$ and $\boldsymbol{\alpha}$ are updated by the optimizer without any projection, and the optimized variables $\bm{\Theta}^*$ and $\bm{P}^*$ of (\textbf{OPT-1}) are obtained from them in the post-processing step through \eqref{eq:phase_forward} and \eqref{eq:power_forward}.}
	\textcolor{black}{Since these mappings map any values of $\boldsymbol{\psi}$ and $\boldsymbol{\alpha}$ to phase shifts and transmit powers that satisfy \eqref{eq:c1p1}, \eqref{eq:c3p1}, and \eqref{eq:c4p1}, the phase shifts and transmit powers corresponding to every intermediate iterate of $\boldsymbol{\psi}$ and $\boldsymbol{\alpha}$ also satisfy these constraints.}
	\textcolor{black}{Hence, the proposed method satisfies these constraints by construction rather than enforcing them through projection, which is its key difference from a direct projected gradient optimization.}
	\textcolor{black}{The latter updates the optimization variables $\bm{\Theta}$ and $\bm{P}$ using gradient-based steps and then projects the updated variables back onto the feasible set defined by these constraints.}
	\textcolor{black}{This projection modifies any update that would otherwise violate the constraints.}
	\textcolor{black}{The two approaches are compared in Section~\ref{subsec:ablation} using the same automatic differentiation framework and optimizer.}
\end{nrem}

\textcolor{black}{Having presented the proposed NR-QoS algorithm, we next discuss its convergence, computational complexity, and implementation considerations.}

\begin{nrem}[Convergence Analysis]\label{remark:convergence}
	Due to the highly nonconvex nature of the neural network parameterization in terms of  $\boldsymbol{\psi}$ and $\boldsymbol{\alpha}$, finding a global optimum of the minimax problem (\textbf{OPT-2}) is analytically intractable.
	Nevertheless, the primal-dual loss function is linear, and thus concave, with respect to the dual variables $\boldsymbol{\mu}$, which places (\textbf{OPT-2}) \textcolor{black}{within} the class of nonconvex-concave minimax problems.
	For such problems, under standard Lipschitz smoothness assumptions \textcolor{black}{and a bounded dual set}, it has been theoretically proven in \cite{lin2020gradient, nouiehed2019solving} that \textcolor{black}{gradient descent and ascent algorithms find} an $\epsilon$-\textcolor{black}{first-order} stationary point (FOSP) when employing tailored learning rates \textcolor{black}{and} multi-step updates\textcolor{black}{, respectively}.
	\textcolor{black}{On any bounded subset of the dual feasible set of (\textbf{OPT-2}), the smoothness assumptions are satisfied by the loss function \eqref{eq:loss_PD}, since $\sigma_k^2>0$ and it depends on $\boldsymbol{\psi}$ only through $e^{j\boldsymbol{\psi}}$ and on $\boldsymbol{\alpha}$ only through the softmax function.}
	\textcolor{black}{Similar guarantees have also been established for stochastic gradient-based optimizers such as Adam in nonconvex minimization \cite{defossez2022simple, zhang2022adam, li2023convergence}, and, under concavity in the dual variables, for their combination with a projected dual ascent step \cite{guo2025unified}, as employed in Algorithm \ref{alg:proposed}.}
	\textcolor{black}{Based on these results, with appropriately tuned learning rates, our proposed NR-QoS framework is expected to approach an FOSP of (\textbf{OPT-2}).}
	\textcolor{black}{The convergence behavior of NR-QoS is presented in Section~\ref{sec:conver}.}
	\textcolor{black}{The connection between such a point and the Karush-Kuhn-Tucker (KKT) conditions of (\textbf{OPT-1}) is established in Lemma~\ref{lemma:kkt}.}
\end{nrem}

\begin{lemma}\label{lemma:kkt}
	\textcolor{black}{Suppose that the solution $(\boldsymbol{\psi}, \boldsymbol{\alpha}, \boldsymbol{\mu})$ generated by Algorithm \ref{alg:proposed} with a stochastic gradient-based optimizer, such as Adam~\cite{Kingma2015}, converges to $(\boldsymbol{\psi}^*, \boldsymbol{\alpha}^*, \boldsymbol{\mu}^*)$ as $t\to\infty$.}
	\textcolor{black}{Then, the phase shifts $\bm{\Theta}^*$ and the power vector $\bm{p}^*$ obtained from $(\boldsymbol{\psi}^*, \boldsymbol{\alpha}^*)$ through \eqref{eq:phase_forward} and \eqref{eq:power_forward} satisfy} the KKT necessary optimality conditions of the original constrained problem (\textbf{OPT-1}).
\end{lemma}
\begin{IEEEproof}
	Please see the detailed proof in Appendix \ref{app:kkt}.
\end{IEEEproof}

\textcolor{black}{We now analyze the computational complexity of the proposed NR-QoS algorithm.}
\textcolor{black}{According to \cite[Section~6.5]{goodfellow2016deep}, the time complexity of updating a parameterized neural network via backpropagation is dominated by the matrix multiplications in the forward propagation, and the backward propagation has the same order.}
\textcolor{black}{In the proposed neural reparameterization architecture, each of the first $L-1$ layers of the phase shift design module multiplies a $K \times N$ matrix by an $N \times N$ transmission matrix, which requires $\mathcal{O}(KN^2)$ operations, while the last layer and the rate calculation module require $\mathcal{O}(K^2N)$ operations for $M=K$.}
\textcolor{black}{Since the transmission matrices are fixed, the backward propagation multiplies the backpropagated gradients by their conjugate transposes at the same cost.}
\textcolor{black}{Hence, the computational complexity for updating the phase shift design module is bounded by $\mathcal{O}(LKN^2)$ per iteration.}
The power allocation module and the corresponding $K$ Lagrangian multipliers associated with the QoS constraints each require $\mathcal{O}(K)$ operations to update.
Let $T$ denote the total number of iterations.
\textcolor{black}{In summary, the overall computational complexity of the proposed algorithm is bounded by $\mathcal{O}(TLKN^2)$.}

{Regarding implementation, the proposed framework can directly handle complex-valued quantities using automatic differentiation.}
{In particular,} $\bm{H}$, $\bm{W}_l$, and $e^{j\boldsymbol{\psi}}$ {in $\mathcal{F}$ need not be split} into separate real and imaginary parts.
Since $\boldsymbol{\psi}$, $\boldsymbol{\alpha}$, and the result of \eqref{eq:loss_PD} are real-valued, it suffices that backpropagation updates $\boldsymbol{\psi}$ and $\boldsymbol{\alpha}$ correctly.
Modern deep learning frameworks, such as PyTorch~\cite{pytorchComplex}, handle this via conjugate Wirtinger derivatives, yielding the required real-valued gradients.

{Finally, we} emphasize that the proposed framework is an optimization method in which the neural network $\mathcal{F}$ serves merely as a computational accelerator for solving (\textbf{OPT-1}), rather than a model that must be trained from a dataset.
For a given channel realization and transmit power budget, the corresponding phase shifts $\bm{\Theta}$ and the power allocation $\bm{P}$ are optimized by Algorithm \ref{alg:proposed}.
Consequently, no training dataset is required, and the conventional training, validation, and testing stages, as well as generalization issues to unseen channel realizations, are not applicable.
Furthermore, the structure of $ \mathcal{F}$ is dictated by the SIM model: the phase shift design module stacks $L$ cascaded layers of dimension  $N$,  governed by the wave-propagation model in \eqref{eq:computeG}, while the power allocation module is a single layer mapping $P_{\max}$  to a feasible power vector.
In other words, the layer count, layer width, and activation function  are  determined by the SIM communication model.

\section{Numerical Results}\label{sec:results}

\subsection{Simulation Setup}\label{subsec:setup}

In the simulations, we consider a downlink SIM-assisted multi-user MISO system where the BS is equipped with a uniform linear array (ULA) of $M = 4$ antennas with half-wavelength spacing. 
The carrier frequency is $28$ GHz.
Each layer of the SIM is modeled as a uniform planar array (UPA) with $N_x \times N_y$ meta-atoms, where square metasurfaces are assumed with $N_x = N_y$ and element spacing $\lambda/2$.
The overall SIM thickness is set to $T_{\mathrm{SIM}} = 5\lambda$, which implies an inter-layer spacing of $T_{\mathrm{SIM}}/L$. The BS height is fixed at $H_{\mathrm{BS}} = 10$ m.
The geometric configuration and inter-layer spatial arrangement of the SIM, as well as the user deployment topology, are adopted consistently with the setup established in \cite{An2023a,An2025}. 
We consider $K=M$ single-antenna users located along a straight line with a separation of $d_{\mathrm{user}} = 10$ m. 
The antenna gains at the BS and the users are set to $5$ dBi and $0$ dBi, respectively. The large-scale fading between the SIM and the $k$-th user is described by a distance-dependent path-loss model $\beta_k = C_0 (d_k/d_0)^{-\bar{n}}$, where \textcolor{black}{$d_k$ is the distance between the SIM and the $k$-th user,} $d_0 = 1~{\rm m}$ is the reference distance, $C_0 = (\lambda/(4\pi d_0))^2$ denotes the free-space path loss at distance $d_0$, and the path-loss exponent is chosen as $\bar{n} = 3.5$ \cite{Rappaport2015}. 
We consider an indoor millimeter-wave (mmWave) scenario with a noise power spectral density of $-174$ dBm/Hz over a 10 MHz system bandwidth, resulting in a noise variance of $\sigma_k^2 = -104$ dBm for all users.
Unless otherwise specified in the figures or discussions, the default simulation parameters are summarized as follows:
the SIM is configured with $L = 6$ layers and $N = 49$ meta-atoms per layer (i.e., $N_x = N_y = 7$), the maximum transmit power at the BS is set to $P_{\max} = 15$ dBm, and all users are subject to the same minimum rate requirement, i.e., $R_{k}^{\min} = 1~{\rm bps/Hz},  \forall 1\leq k \leq K$.

To comprehensively evaluate the performance of our proposed approach, we compare it against several benchmark schemes. 
While all the considered baselines share the common objective of maximizing the achievable sum rate, they differ significantly in their constraint configurations and underlying optimization methodologies.
These benchmarks include:
\begin{itemize}
	\item \textbf{Neural reparameterization without QoS constraints (NR-w/o-QoS)}:
	This method utilizes our proposed neural reparameterization framework to jointly design the phase shift control and the power allocation. 
	The primary distinction is that this scheme does not consider the minimum data rate constraint. 
	Therefore, the parameters of the neural network are directly updated according to the loss function $ \mathcal{L}_{\mathrm {NR\text{-}w\text{/}o\text{-}QoS}}(\boldsymbol{\psi}, \boldsymbol{\alpha} ) = -\sum_{k=1}^{K}  r_k(\boldsymbol{\psi}, \boldsymbol{\alpha})$.
	\item \textbf{AO}:
	The AO algorithm proposed in \cite{An2023} serves as a representative baseline. 
	Similar to the NR-w/o-QoS scheme, this method focuses strictly on maximizing the achievable sum rate without imposing any minimum individual QoS constraints.
	The algorithm decomposes the joint optimization into two separate subproblems: a modified iterative water-filling procedure for power allocation and a gradient-ascent procedure for phase shift control.
	\item \textcolor{black}{\textbf{Augmented Lagrangian method for QoS constraints (AL-QoS)}:}
	\textcolor{black}{This benchmark is based on the augmented Lagrangian of (\textbf{OPT-1}), in which a Lagrangian  multiplier is introduced for the QoS constraint \eqref{eq:c2p1} of each user.}
	\textcolor{black}{The transmit power allocation is updated by projected gradient ascent on the augmented Lagrangian.}
	\textcolor{black}{The phase shifts are updated by gradient ascent with backtracking line search in a similar way to the AO algorithm in \cite{An2023}, but using the gradient of the augmented Lagrangian, which incorporates the QoS constraints through the Lagrangian multipliers.}
	\textcolor{black}{These two updates are performed alternately, followed by the update of the multipliers.}
	\item \textcolor{black}{\textbf{VD-BSUM}:}
	\textcolor{black}{The VD-BSUM algorithm proposed in \cite{Zhang2026} is a recent numerical optimization method that also addresses the QoS-constrained sum rate maximization problem (\textbf{OPT-1}) in SIM-assisted systems.}
	\textcolor{black}{It first reformulates the problem via the weighted minimum mean square error (WMMSE) method, and then sequentially updates the power allocation and the phase shifts of each metasurface layer through closed-form iterations.}
\end{itemize}

\textcolor{black}{Unless otherwise specified,} all simulation results are averaged over \textcolor{black}{300} independent channel realizations\textcolor{black}{, where the average sum rate of the QoS-aware schemes is calculated over the channel realizations in which a feasible solution is obtained}.
The simulations are executed on a \textcolor{black}{mobile} workstation equipped with an Intel \textcolor{black}{Core i9-14900HX} central processing unit (CPU) and \textcolor{black}{64 GB random-access memory (RAM)}.
The proposed framework is implemented in PyTorch \textcolor{black}{2.3.0} \cite{pytorch}, the maximum number of iterations for our algorithm is set to \textcolor{black}{$T=100$}, and the parameters \textcolor{black}{$\boldsymbol{\psi}$ and $\boldsymbol{\alpha}$} are updated by the Adam optimizer with learning \textcolor{black}{rates $0.45$ and $0.0045$, respectively} \cite{Kingma2015}.
For the baseline AO algorithm proposed in~\cite{An2023}, all parameters follow the official implementation released by the authors~\cite{AnGithub}, with the maximum number\textcolor{black}{s} of \textcolor{black}{iterations for the AO algorithm, the iterative water-filling, and the backtracking line search} set to \textcolor{black}{100, as specified in~\cite{An2025}}.
\textcolor{black}{The maximum number of outer iterations of AL-QoS and VD-BSUM is also set to 100.}
For a fair computational comparison, we \textcolor{black}{implement the AO, AL-QoS, and VD-BSUM benchmarks in both NumPy and PyTorch} and run all competing methods on the CPU under identical hardware settings, starting from the same \textcolor{black}{initial point for each channel realization, where the phase shifts $\bm{\Theta}^{(0)}$ are} randomly generated \textcolor{black}{and the transmit power is equally allocated among the users}.

\subsection{Convergence Performance}\label{sec:conver}

\begin{figure*}[!t]
	\centering
	\subfloat[{\textcolor{black}{Sum rate.}}]{\includegraphics[width=0.32\textwidth]{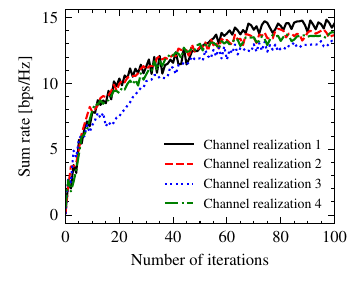}\label{fig:cover_rate}}
	\hfil
	\subfloat[{\textcolor{black}{Minimum user rate.}}]{\includegraphics[width=0.32\textwidth]{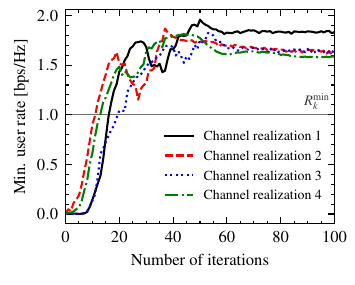}\label{fig:cover_minrate}}
	\hfil
	\subfloat[{\textcolor{black}{Dual variables.}}]{\includegraphics[width=0.32\textwidth]{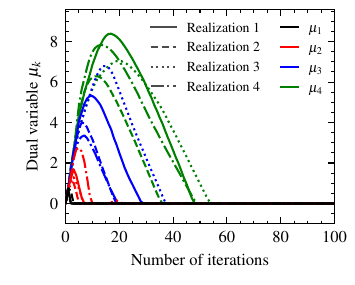}\label{fig:cover_dual}}
	\caption{\textcolor{black}{Convergence behavior of the proposed NR-QoS algorithm for four channel realizations with $L=6$, $N=49$, $P_{\max}=15$~dBm, and $R_k^{\min}=1$~bps/Hz.}}
	\label{fig:cover}
\end{figure*}

We first examine the convergence behavior of the proposed NR-QoS algorithm.
\textcolor{black}{\figref{fig:cover} depicts the evolution of the sum rate, the minimum user rate, and the dual variables with respect to the number of iterations for $L=6$ and $N=49$.}
The results are obtained from four independent channel realizations.
\textcolor{black}{It can be observed from \figref{fig:cover}\subref{fig:cover_rate} that the sum rate does not increase in an absolute strictly monotonic manner as the number of iterations grows.}
This behavior is expected since the proposed framework uses a stochastic gradient-based optimizer, which may introduce mild oscillations during the update process.
\textcolor{black}{Nevertheless, the overall trend exhibits relatively stable convergence, where the sum rate increases rapidly during the first $20$ iterations and gradually saturates afterwards.}
\textcolor{black}{As shown in \figref{fig:cover}\subref{fig:cover_minrate}, the minimum user rate exceeds the threshold $R_k^{\min}$ within $30$ iterations and remains above it afterwards, i.e., the QoS constraints of all users are satisfied.}
\textcolor{black}{Correspondingly, the dual variables in \figref{fig:cover}\subref{fig:cover_dual} increase while the QoS constraints are violated and decrease to zero after the constraints are satisfied, which agrees with the complementary slackness condition.}

\subsection{Effect of the Number of Metasurface Layers}
\begin{figure}[!htbp]
	\center
	\includegraphics[width=1\columnwidth]{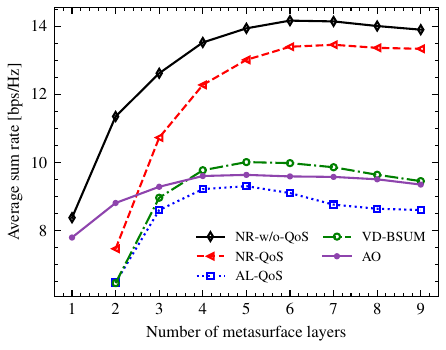}
	\caption{Average sum rate versus the number of metasurface layers $L$.}
	\label{fig:layer}
\end{figure}

\figref{fig:layer} illustrates the average sum rate as a function of the number of metasurface layers $L$.
\textcolor{black}{The lack of plotted points for the proposed NR-QoS algorithm as well as the AL-QoS and VD-BSUM schemes at $L=1$ indicates that no feasible solution is found, as a single layer is unable to meet the considered QoS requirements.}
It is observed from \figref{fig:layer} that increasing the number of cascaded SIM layers initially leads to a noticeable improvement in the achievable sum rate for all considered schemes.
\textcolor{black}{Under the same QoS constraints, the proposed NR-QoS algorithm achieves 40.5\% and 30.2\% sum rate improvements over the AL-QoS and VD-BSUM schemes, respectively, on average over $2 \leq L \leq 9$.}
\textcolor{black}{These improvements indicate that embedding the cascaded SIM propagation model into a structured neural network, whose parameters are jointly updated, enables more effective exploitation of the layered EM structure than relying on the alternating updates of the benchmark schemes.}
\textcolor{black}{Although all methods are optimized from the same initial point, the sum rate of NR-QoS remains almost unchanged once $L > 5$, whereas the AL-QoS and VD-BSUM schemes suffer a noticeable performance drop.}
\textcolor{black}{This performance contrast indicates that the degradation mainly arises from the intrinsic difficulty encountered by these benchmarks in solving the increasingly high-dimensional nonconvex optimization problem.}
In fact, even our proposed NR-QoS scheme outperforms the AO scheme (that does not consider QoS constraints) as the number of metasurface layers increases, highlighting the remarkable performance of the proposed neural reparameterization framework.

\subsection{Effect of the Number of Meta-Atoms per Layer}

\begin{figure}[!htbp]
	\center
	\includegraphics[width=1\columnwidth]{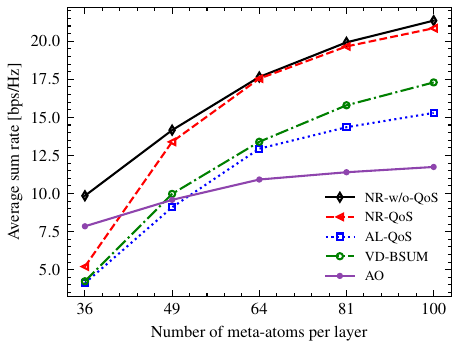}
	\caption{Average sum rate versus the number of meta-atoms per layer $N$.}
	\label{fig:atoms}
\end{figure}

\figref{fig:atoms} shows the average sum rate versus the number of meta-atoms per layer.
It is clear that the average sum rate increases as the number of meta-atoms grows for all schemes.
\textcolor{black}{At $N = 100$, the proposed NR-QoS algorithm outperforms the AO benchmark by 77.5\%, despite operating under additional QoS constraints.}
\textcolor{black}{Compared with the AL-QoS and VD-BSUM schemes, which are subject to the same QoS constraints, the corresponding improvements are 36.3\% and 20.6\%, respectively.}
\textcolor{black}{This observation suggests that directly embedding the physical propagation model into the neural architecture is more effective than relying on the alternating updates of the benchmark schemes, which is consistent with the trend shown in \figref{fig:layer}.}
\textcolor{black}{When $N \geq 64$, the sum rate of NR-QoS is within 2.4\% of that of NR-w/o-QoS.}
\textcolor{black}{Hence, the proposed NR-QoS algorithm effectively exploits the additional spatial DoF brought by larger metasurface arrays to satisfy the QoS constraints with a negligible sum rate loss.}
\textcolor{black}{This reveals the effectiveness of our proposed neural reparameterization framework for joint power allocation and SIM phase shift design optimization under QoS constraints.}

\subsection{Effect of the Maximum Transmit Power at the BS}

\begin{figure}[!htbp]
	\center
	\includegraphics[width=1\columnwidth]{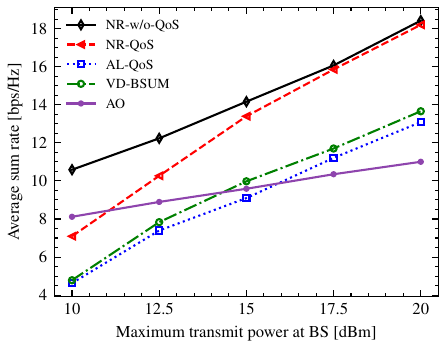}
	\caption{Average sum rate versus the maximum transmit power $P_{\max}$.}
	\label{fig:power}
\end{figure}

In \figref{fig:power}, we plot the impact of the maximum transmit power budget $P_{\max}$ on the achievable average sum rate.
Although the average sum rate increases monotonically with $P_{\max}$, the rate growth behavior differs significantly across the considered schemes.
\textcolor{black}{For instance, the proposed NR-QoS algorithm exhibits the largest increase in average sum rate as the power budget grows, whereas the AO method exhibits the smallest increase compared with all other methods.}
\textcolor{black}{This reveals the limitation of AO in designing an effective phase shift policy to fully utilize a higher transmit power budget.}
\textcolor{black}{When $P_{\max} \geq 17.5$ dBm, the proposed NR-QoS algorithm achieves at least a 53.1\% average performance improvement over AO.}
\textcolor{black}{Moreover, NR-QoS outperforms the AL-QoS and VD-BSUM schemes by at least 38.8\% and 31.0\%, respectively, over the considered range of $P_{\max}$.}
\textcolor{black}{Similar to the observations in \figref{fig:atoms}, the sum rate of NR-QoS approaches that of NR-w/o-QoS as the power budget grows, i.e., stays within 1.4\% when $P_{\max} \geq 17.5$ dBm.}
\textcolor{black}{On the other hand, when the transmit power budget is low, more power has to be allocated to the users with weak channels in order to satisfy their QoS requirements, resulting in a larger sum rate difference.}

\subsection{Effect of the Minimum Data Rate Threshold}
\begin{figure}[!htbp]
	\center
	\includegraphics[width=1\columnwidth]{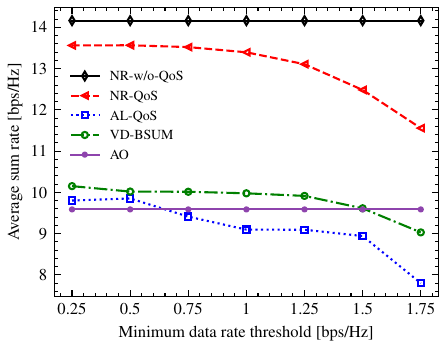}
	\caption{Average sum rate versus the minimum data rate threshold $R^{\min}_k$.}
	\label{fig:rmin}
\end{figure}

\begin{figure}[!htbp]
	\center
	\includegraphics[width=1\columnwidth]{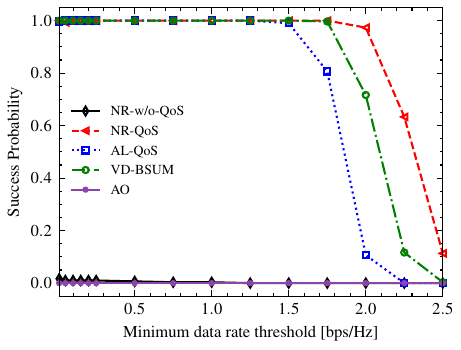}
	\caption{Success probability versus the minimum data rate threshold $R^{\min}_k$.}
	\label{fig:succprob}
\end{figure}

\figref{fig:rmin} and \figref{fig:succprob} depict the effect of the minimum data rate threshold on the average sum rate and the success probability, respectively.
\textcolor{black}{The success probability is defined as the fraction of the number of feasible solutions found over the 300 considered channel realizations, where a higher value indicates better performance.}
\textcolor{black}{The results of the NR-w/o-QoS and AO schemes were obtained without considering QoS constraints, so their sum rate performance shown in \figref{fig:rmin} is independent of $R_k^{\min}$.}
As expected, increasing the minimum data rate requirement renders the QoS-aware schemes more restrictive in terms of average sum rate.
\textcolor{black}{Nevertheless, the proposed NR-QoS scheme outperforms the AL-QoS and VD-BSUM schemes by at least 37.6\% and 27.9\%, respectively, for all the thresholds considered in \figref{fig:rmin}.}
\textcolor{black}{In contrast to the AL-QoS and VD-BSUM schemes, whose average sum rates become smaller than that of the AO baseline when $R_k^{\min}$ exceeds a certain threshold, NR-QoS outperforms AO for all the considered thresholds.}
\textcolor{black}{Even at $R_k^{\min} = 1.75$ bps/Hz, the average sum rate of NR-QoS is still 20.6\% higher than that of AO, although the former has to satisfy the QoS constraints whereas the latter does not.}
\textcolor{black}{Moreover, it is worth noting that AO and NR-w/o-QoS maintain a nearly zero success probability as shown in \figref{fig:succprob}, since the minimum achievable data rate of at least one user in the system does not meet the minimum rate threshold and is in fact observed to be on the order of $10^{-9}$ for AO.}
\textcolor{black}{In contrast, the proposed NR-QoS algorithm achieves a success probability close to one up to $R_k^{\min} = 2$ bps/Hz, whereas the success probabilities of the AL-QoS and VD-BSUM schemes drop rapidly once $R_k^{\min}$ exceeds 1.5 bps/Hz and 1.75 bps/Hz, respectively.}
\textcolor{black}{As $R_k^{\min}$ further increases, NR-QoS still maintains a significantly higher success probability than these two schemes.}

\subsection{Computational Runtime Analysis}\label{subsec:runtime}

\begin{table}[!t]
	\caption{\textcolor{black}{Average Computational Time (Seconds) Under Different $L$ and $N$}}
	\label{tab:complexity}
	\centering
	\setlength{\tabcolsep}{2.5pt}
	\color{black}
	\begin{tabular}{l|cccc|ccc}
		\hline
		\hline
		& \multicolumn{4}{c|}{$L$ ($N=49$)} & \multicolumn{3}{c}{$N$ ($L=6$)} \\
		\cline{2-8}
		Method & 1 & 3 & 6 & 9 & 36 & 49 & 100 \\
		\hline
		\hline
		AO (NumPy) & 0.316 & 0.781 & 0.903 & 0.947 & 0.883 & 0.903 & 1.016 \\
		\hline
		AO (PyTorch) & 0.340 & 0.841 & 0.994 & 1.041 & 0.962 & 0.994 & 1.089 \\
		\hline
		AL-QoS (NumPy) & 0.105 & 0.355 & 0.475 & 0.581 & 0.229 & 0.475 & 0.731 \\
		\hline
		AL-QoS (PyTorch) & 0.179 & 0.534 & 0.696 & 0.818 & 0.374 & 0.696 & 0.795 \\
		\hline
		VD-BSUM (NumPy) & 0.747 & 1.337 & 2.409 & 3.546 & 2.118 & 2.409 & 3.626 \\
		\hline
		VD-BSUM (PyTorch) & 1.439 & 2.160 & 3.692 & 5.490 & 3.519 & 3.692 & 4.932 \\
		\hline
		NR-w/o-QoS (PyTorch) & 0.062 & 0.071 & 0.083 & 0.097 & 0.082 & 0.083 & 0.095 \\
		\hline
		NR-QoS (PyTorch) & 0.067 & 0.077 & 0.088 & 0.103 & 0.088 & 0.088 & 0.100 \\
		\hline
		\hline
	\end{tabular}
\end{table}

\textcolor{black}{Table~\ref{tab:complexity} summarizes the average computational time under varying numbers of SIM layers $L$ and meta-atoms per layer $N$.}
\textcolor{black}{The average computational time is obtained by averaging the total execution  time over 100 independent channel realizations, where each method runs on a single CPU thread.}
\textcolor{black}{To exclude the influence of the implementation framework, the runtime of the AO, AL-QoS, and VD-BSUM benchmarks is reported for both their NumPy and PyTorch implementations.}
\textcolor{black}{Regardless of the implementation framework, the proposed NR-QoS algorithm requires a lower runtime than all three benchmarks in every considered configuration.}
\textcolor{black}{For instance, when $L=6$ and $N=49$, NR-QoS is approximately 10, 5, and 27 times faster than the NumPy implementations of AO, AL-QoS, and VD-BSUM, respectively.}
\textcolor{black}{Moreover, the runtime of the AL-QoS and VD-BSUM benchmarks increases rapidly as either $L$ or $N$ increases, which reflects their sensitivity to the dimensional expansion of the optimization variables.}
\textcolor{black}{This behavior is expected since these benchmarks repeatedly solve high-dimensional subproblems whose complexity increases with the total number of phase-shift parameters.}
In contrast, the proposed neural reparameterization framework exhibits a significantly milder growth pattern with respect to both $L$ and $N$.
\textcolor{black}{Compared with NR-w/o-QoS, NR-QoS requires only a slightly higher runtime due to the update of the Lagrangian multipliers.}
\textcolor{black}{The results shown in Table~\ref{tab:complexity} confirm that the proposed neural reparameterization framework preserves computational efficiency even in large-scale SIM configurations and requires a lower runtime than all the benchmark schemes, which makes it a feasible solution for implementing wave-domain signal processing in real-time wireless communication systems.}

\subsection{\textcolor{black}{Ablation Study and Comparison with DRL}}\label{subsec:ablation}

\begin{figure}[!htbp]
	\center
	\includegraphics[width=1\columnwidth]{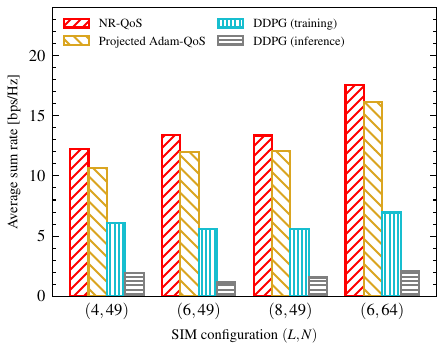}
	\caption{\textcolor{black}{Average sum rate of NR-QoS, Projected Adam-QoS, and DDPG under different SIM configurations $(L, N)$.}}
	\label{fig:ablation}
\end{figure}

\textcolor{black}{Since the proposed framework updates the learnable neural parameters $\boldsymbol{\psi}$ and $\boldsymbol{\alpha}$ with the Adam optimizer, it is necessary to examine whether its performance gain stems from the optimizer alone.}
\textcolor{black}{To this end, we consider a direct projected gradient optimization scheme, referred to as Projected Adam-QoS, which applies automatic differentiation and the Adam optimizer directly to the original variables $\bm{\Theta}$ and $\bm{P}$ of (\textbf{OPT-1}).}
\textcolor{black}{This scheme adopts the same loss function, dual update, learning rates, and number of iterations as NR-QoS.}
\textcolor{black}{After each update, the phase shifts and the transmit powers are projected onto the feasible set defined by \eqref{eq:c1p1}, \eqref{eq:c3p1}, and \eqref{eq:c4p1}.}
\textcolor{black}{In addition, we compare the proposed framework with a DRL benchmark proposed in \cite{Liu2025b}, which maximizes the sum rate without QoS constraints by the deep deterministic policy gradient (DDPG) algorithm \cite{Lillicrap2016}.}
\textcolor{black}{In our implementation, each agent is trained on the first 100 of the 300 channel realizations, which are processed in parallel, for 50 episodes of 26,000 steps with the hyperparameters in \cite{Liu2025b}.}

\textcolor{black}{Fig.~\ref{fig:ablation} compares the average sum rate of these schemes under four SIM configurations. Two results are reported for the DDPG scheme. DDPG (training) is the average of the best sum rates found during training for the 100 training channel realizations, whereas DDPG (inference) is the average sum rate obtained by the trained DRL agent over the 300 channel realizations. It is observed that NR-QoS outperforms Projected Adam-QoS by 8.6\% to 14.8\% in all the considered configurations, although both schemes employ the same Adam optimizer. This demonstrates that the performance improvement cannot be attributed solely to the choice of the automatic differentiation method and the Adam optimizer, but also comes from the proposed reparameterization that allows the phase and transmit power constraints to be satisfied by construction, so that the Adam updates are not altered by any projection.
}

\textcolor{black}{
	In comparison with the DRL-based scheme, NR-QoS consistently achieves a significantly higher average sum rate across the considered configurations. The performance gap between DDPG (training) and DDPG (inference) highlights the difference between the best sum rates achieved during training and the performance of the trained agent during inference, which includes 200 channel realizations unseen during training, suggesting potential limitations in the generalization of the learned policy \cite{Cobbe2019}. It should also be noted that the DDPG scheme in \cite{Liu2025b} does not explicitly consider individual user QoS constraints, whereas NR-QoS incorporates these constraints through Lagrangian multipliers.
}

\textcolor{black}{
	Moreover, training one DDPG agent, whose actor and critic networks contain 3.6 to 6.9 million parameters in our implementation, requires $1.3\times10^{6}$ updates of these networks and takes about six hours on an NVIDIA RTX 3500 Ada graphics processing unit (GPU). In addition, since the dimensions of these networks depend on $L$, $N$, and $K$, a separate DDPG agent has to be trained for each SIM configuration, which incurs a substantial training overhead. In contrast, NR-QoS requires no offline training and solves (\textbf{OPT-1}) for each channel realization within $T = 100$ iterations over only $LN+2K$ variables, which take at most 0.103 seconds on a single CPU thread for the configurations in Table~\ref{tab:complexity}. As an illustration of the difference in computational requirements, during the time needed to train one DDPG agent on the specified GPU, NR-QoS can solve (\textbf{OPT-1}) for more than $2\times10^{5}$ channel realizations on a single CPU thread.
}

\textcolor{black}{
	Overall, these results demonstrate the effectiveness of the proposed framework in achieving higher sum rates than the considered direct gradient-based optimization and DRL baselines, while avoiding the offline training overhead associated with DRL.
}

\section{Conclusion}\label{sec:conclusion}

This paper presented a neural reparameterization framework for SIM-assisted multi-user precoding to address the severe computational overhead and scalability limitations of conventional numerical optimization methods. 
The proposed approach embeds the cascaded EM wave-propagation structure of the SIM directly into a differentiable neural network architecture, where the SIM phase shifts and transmit power allocation are obtained from neural parameters that are jointly optimized via a primal-dual learning procedure to maximize the sum rate under QoS constraints. 
Distinct from black-box learning, the framework uses the neural network to reparameterize and optimize the precoding variables for each channel realization, rather than to learn a data-driven channel-to-precoding mapping, thereby requiring no offline training and avoiding the associated generalization issues.
Extensive simulations verified that the proposed method achieved improved sum rate performance while simultaneously delivering a significant reduction in computational runtime compared with traditional numerical optimization-based algorithms.
This work therefore offers a promising path toward deploying real-time SIM-assisted wireless communication systems.

\textcolor{black}{The proposed framework offers several promising directions for future research.}
\textcolor{black}{In particular, extending the proposed framework to SIMs with discrete phase shifts and developing precoding designs that are robust to imperfect CSI and hardware imperfections are of practical importance.}
\textcolor{black}{Moreover, incorporating more physically consistent SIM models that account for the mutual coupling among the meta-atoms is another important direction for future research.}

\begin{appendices}
	\section{Proof of Lemma \ref{lemma:kkt}}\label{app:kkt}
	{\color{black}
	With a proper learning rate $\eta_{\psi} > 0$, the Adam update of $\boldsymbol{\psi}$ converges to $-\eta_{\psi}\, \bm{g}^*/(|\bm{g}^*| + \delta)$, where $\bm{g}^* \triangleq \nabla_{\boldsymbol{\psi}} \mathcal{L}(\boldsymbol{\psi}^*, \boldsymbol{\alpha}^*, \boldsymbol{\mu}^*)$, $\delta > 0$ is the stabilizing constant of Adam~\cite{Kingma2015}.
	If $\boldsymbol{\psi}$ converges, its limit is a fixed point of the Adam iteration, at which the update $-\eta_{\psi}\, \bm{g}^*/(|\bm{g}^*| + \delta)$ is zero, and thus we can have $\bm{g}^* = \bm{0}$~\cite{Bock2019}.
	Applying the same argument to $\boldsymbol{\alpha}$, we can have
	\begin{align}
		\nabla_{\boldsymbol{\psi}} \mathcal{L}(\boldsymbol{\psi}^*, \boldsymbol{\alpha}^*, \boldsymbol{\mu}^*) &= \bm{0}, \label{eq:kkt_stat1} \\
		\nabla_{\boldsymbol{\alpha}} \mathcal{L}(\boldsymbol{\psi}^*, \boldsymbol{\alpha}^*, \boldsymbol{\mu}^*) &= \bm{0}. \label{eq:kkt_stat2}
	\end{align}

	Next, the projection in \eqref{eq:updates} gives $\mu_k \geq 0$ throughout the iterations.
	Moreover, $R_k^{\min} - r_k(\boldsymbol{\psi}, \boldsymbol{\alpha})$ converges to $c_k^* \triangleq R_k^{\min} - r_k(\boldsymbol{\psi}^*, \boldsymbol{\alpha}^*) = R_k^{\min} - r_k(\bm{\Theta}^*, \bm{p}^*)$.
	If $c_k^* > 0$, then $R_k^{\min} - r_k(\boldsymbol{\psi}, \boldsymbol{\alpha}) \geq c_k^*/2$ for all sufficiently large $t$, and each update in \eqref{eq:updates} then increases $\mu_k$ by at least $\eta_{\mu}^{(t)} c_k^*/2$.
	If $c_k^* < 0$ with $\mu_k^* > 0$, the projection in \eqref{eq:updates} is inactive for all sufficiently large $t$, and each update then decreases $\mu_k$ by at least $\eta_{\mu}^{(t)} |c_k^*|/2$.
	Since $\sum_t \eta_{\mu}^{(t)} = \infty$, both cases contradict the convergence of $\mu_k$.
	Therefore, $\mu_k^* \geq 0$, $c_k^* \leq 0$, and $\mu_k^* c_k^* = 0$ for all $k$, i.e., the KKT conditions associated with \eqref{eq:c2p1} hold.

	Let $\boldsymbol{\mu} \succeq \bm{0}$ and $\nu \geq 0$ denote the Lagrangian multipliers of \eqref{eq:c2p1} and \eqref{eq:c1p1}, respectively.
	The Lagrangian function of (\textbf{OPT-1}) associated with \eqref{eq:c1p1} and \eqref{eq:c2p1} is \cite[Section~5.1.1]{Boyd2004}
	\begin{align}
		\mathcal{L}_{\mathrm{o}}(\bm{\Theta}, \bm{p}, \boldsymbol{\mu}, \nu) = & \sum_{k=1}^{K} r_k + \sum_{k=1}^{K} \mu_k \big(r_k - R_k^{\min}\big) \notag\\
		& + \nu \Big(P_{\max} - \sum_{k=1}^{K} p_k\Big). \label{eq:kkt_Lo}
	\end{align}

	By \eqref{eq:phase_forward}, each phase shift in $\bm{\Theta}^*$ differs from the corresponding entry of $\boldsymbol{\psi}^*$ by an integer multiple of $2\pi$.
	Since $\mathcal{L}_{\mathrm{o}}$ depends on each $\theta_n^l$ only through $e^{j\theta_n^l}$, $\nabla_{\bm{\Theta}} \mathcal{L}_{\mathrm{o}}$ is $2\pi$-periodic in every $\theta_n^l$ and thus takes the same value at $\boldsymbol{\psi}^*$ as at $\bm{\Theta}^*$.
	Then, \eqref{eq:loss_PD} and \eqref{eq:kkt_Lo} show that \eqref{eq:kkt_stat1} is equivalent to
	\begin{align}
		\nabla_{\bm{\Theta}} \mathcal{L}_{\mathrm{o}}(\bm{\Theta}^*, \bm{p}^*, \boldsymbol{\mu}^*, \nu^*) = \bm{0}, \label{eq:kkt_theta}
	\end{align}
	where $\nu^*$ denotes the Lagrangian multiplier associated with \eqref{eq:c1p1} at $(\bm{\Theta}^*, \bm{p}^*)$.

	Since the last term of \eqref{eq:kkt_Lo} is zero for the transmit powers given by \eqref{eq:power_forward}, $\mathcal{L}(\boldsymbol{\psi}, \boldsymbol{\alpha}, \boldsymbol{\mu}) = -\mathcal{L}_{\mathrm{o}}(\bm{\Theta}, \bm{p}, \boldsymbol{\mu}, \nu)$ holds for every $\nu$ when $(\bm{\Theta}, \bm{p})$ is obtained from $(\boldsymbol{\psi}, \boldsymbol{\alpha})$ as in Lemma~\ref{lemma:kkt}.
	Applying the chain rule to \eqref{eq:loss_PD} with \eqref{eq:power_forward} and \eqref{eq:kkt_Lo}, we can have that \eqref{eq:kkt_stat2} is equivalent to
	\begin{align}
		& \nabla_{\bm{p}} \mathcal{L}_{\mathrm{o}}(\bm{\Theta}^*, \bm{p}^*, \boldsymbol{\mu}^*, \nu^*) \notag\\
		& = \Big( \frac{1}{P_{\max}}\sum_{k=1}^{K}(1+\mu_k^*)\, (\bm{p}^*)^{\top} \nabla_{\bm{p}} r_k(\bm{\Theta}^*, \bm{p}^*) - \nu^* \Big) \mathbf{1}_K. \label{eq:kkt_normal}
	\end{align}

	The stationarity condition of (\textbf{OPT-1}) with respect to $\bm{p}$, i.e., $\nabla_{\bm{p}} \mathcal{L}_{\mathrm{o}}(\bm{\Theta}^*, \bm{p}^*, \boldsymbol{\mu}^*, \nu^*) = \bm{0}$, can be obtained by setting
	\begin{align}
		\nu^* = \frac{1}{P_{\max}}\sum_{k=1}^{K}(1+\mu_k^*)\, (\bm{p}^*)^{\top} \nabla_{\bm{p}} r_k(\bm{\Theta}^*, \bm{p}^*). \label{eq:kkt_nu}
	\end{align}

	Each term $(\bm{p}^*)^{\top} \nabla_{\bm{p}} r_k(\bm{\Theta}^*, \bm{p}^*)$ in \eqref{eq:kkt_nu} is the derivative of $r_k(\bm{\Theta}^*, \tau \bm{p}^*)$ with respect to $\tau$ at $\tau = 1$.
	By \eqref{eq:SINR} and \eqref{eq:user_rate}, a common scaling of all transmit powers by $\tau \geq 1$ does not decrease any SINR.
	This derivative is thus nonnegative, which gives $\nu^* \geq 0$.

	In summary, the Lagrangian multipliers $\boldsymbol{\mu}^* \succeq \bm{0}$ and $\nu^* \geq 0$ satisfy the corresponding complementary slackness conditions.
	The stationarity conditions hold by \eqref{eq:kkt_theta} and by \eqref{eq:kkt_normal} with \eqref{eq:kkt_nu}.
	Since \eqref{eq:c3p1} and \eqref{eq:c4p1} are satisfied by construction, their KKT conditions hold with zero Lagrangian multipliers.
	Accordingly, $(\bm{\Theta}^*, \bm{p}^*)$ obtained from the limit $(\boldsymbol{\psi}^*, \boldsymbol{\alpha}^*)$ of Algorithm~\ref{alg:proposed} satisfies the KKT conditions of (\textbf{OPT-1}), which completes the proof.
	}
\end{appendices}

\bibliographystyle{IEEEtran}
\bibliography{IEEEabrv,hokie-HD}
\end{document}